\documentclass[10pt]{article}
\usepackage{euscript,amssymb}

\font\msbm=msbm10
\def\RR{\hbox{\msbm R}}

\def\ZZ{\hbox{\msbm Z}}

\catcode`\@=11
\def\lesssim{\mathrel{\mathpalette\vereq<}}
\def\vereq#1#2{\lower3pt\vbox{\baselineskip1.5pt \lineskip1.5pt
\ialign{$\m@th#1\hfill##\hfil$\crcr#2\crcr\sim\crcr}}}
\def\gtrsim{\mathrel{\mathpalette\vereq>}}

\def\Let@{\relax\iffalse{\fi\let\\=\cr\iffalse}\fi}
\def\vspace@{\def\vspace##1{\crcr\noalign{\vskip##1\relax}}}
\def\multilimits@{\bgroup\vspace@\Let@
 \baselineskip\fontdimen10 \scriptfont\tw@
 \advance\baselineskip\fontdimen12 \scriptfont\tw@
 \lineskip\thr@@\fontdimen8 \scriptfont\thr@@
 \lineskiplimit\lineskip
 \vbox\bgroup\ialign\bgroup\hfil$\m@th\scriptstyle{##}$\hfil\crcr}
\def\Sb{_\multilimits@}
\def\endSb{\crcr\egroup\egroup\egroup}
\def\Sp{^\multilimits@}

\newcommand{\be}[1]{\begin{equation}\label{#1}}
\newcommand{\ee}{\end{equation}}
\newcommand{\ba}[1]{\begin{eqnarray}\label{#1}}
\newcommand{\ea}{\end{eqnarray}}
\newcommand{\rf}[1]{(\ref{#1})}
\newcommand{\nn}{\nonumber}
\newcommand{\const}{\mbox{\rm const}}

\newcommand{\sign}{ \mbox{\rm sign} }

\begin{document}

\author{U. G\"unther\dag\thanks{e-mail: u.guenther@htw-zittau.de},
A. Zhuk\ddag\S\thanks{e-mail: zhuk@paco.odessa.ua}\\[2ex]
\dag Gravitationsprojekt, Mathematische Physik I,\\
Institut f\"ur Mathematik,
Universit\"at Potsdam,\\
Am Neuen Palais 10, PF 601553, D-14415 Potsdam, Germany\\[1ex]
\ddag Department of Physics, University of Odessa, \\
2 Petra Velikogo St., Odessa 65100, Ukraine \\[1ex]
\S
Max-Planck-Institut f\"ur Gravitationsphysik, \\
Albert-Einstein-Institut,\\
Am M\"uhlenberg 1,  D-14476 Golm bei Potsdam, Germany}
\title{Stabilization of internal spaces
in multidimensional cosmology}

\date{29.02.2000}

\maketitle

\abstract{%
Effective 4-dimensional theories are investigated which were
obtained under dimensional reduction of multidimensional cosmological
models with a minimal coupled  scalar field as a matter source.
Conditions for the
internal space stabilization are considered and the possibility
for inflation in the external
space is discussed. The electroweak as well as the Planck
fundamental scale approaches are investigated and compared with each other.
It is shown that there
exists a rescaling for the effective
cosmological constant as well as for gravitational exciton masses in the
different approaches.
}

\bigskip

\hspace*{0.950cm} PACS number(s): 04.50.+h, 98.80.Hw
%
%
%
%
\section{Introduction}
\setcounter{equation}{0}


Stabilization of additional dimensions near their present day values
(dilaton/geometrical moduli stabilization)
is
one of the main problems for any multidimensional theory
because a dynamical
behavior of the internal spaces results in a variation of the fundamental
physical constants. Observations show that internal spaces should be
static or nearly static at least from the time of
recombination (in some papers arguments are given in favor of
 the assumption that
variations of the
fundamental constants are absent
from the time of primordial nucleosynthesis \cite{Kolb}).

Observations further indicated that Standard Model (SM) matter cannot
propagate a large distance in extra dimensions. This allowed for two
classes of model building implications:

The first class consists of models
 with
 extra spacetime dimensions compactified at
scales less the Fermi length $L_F \sim 10^{-17}\, \mbox{\rm cm}$
as characteristic scale
of the experimentally tested electroweak interaction $L_F \sim M^{-1}_{EW}
 \sim 1 \, \mbox{\rm TeV}^{-1}$. Up to the early 1990s this
was a standard assumption
in string phenomenology with string scale slightly below
the 4-dimensional Planck scale $M_s \sim 10^{16}$ GeV \cite{iban}.
The question about a concrete mechanism for the stabilization of
the compactification scales
 (moduli stabilization) remained open in this discussion \cite{iban,BBSMS}.

The second class of models starts from the assumptions that
observable SM matter is confined to a 3-brane located
in a higher dimensional bulk spacetime and that gravitational interactions
can propagate
in the whole bulk spacetime provided that
a mechanism exists which ensures usual
Newton's $r^{-2}$ law at distances $\gtrsim 1$ cm accessable
to present gravitational tests.
The thickness of the 3-brane in this case should be
of order of the Fermi length $L_F$.
The additional bulk dimensions can be compactified or
non-compact.

Historically, the first proposal for an interpretation
of our appearantly 4-dimensional Universe as a submanifold embedded into a
non-compact higher dimensional bulk space dates back to the 1983 work of
Rubakov and Shaposhnikov \cite{RSh} and Akama \cite{Akama}
(still without accounting for
gravitational interactions)
and Visser's consideration from 1985
 \cite{Vi} (studying the
localization/trapping of particles via gravity to
a 4-dimensional submanifold of a 5-dimensional "real" world).

Within the framework of superstring theory/M-theory new arguments have been
given
for a
selfcontent embedding of the 4-dimensional
$SU(3)\times SU(2)\times U(1)$ Standard Model
of strong and electroweak interactions into a fundamentally
higher dimensional spacetime manifold. For example, in Ho\v rava-Witten
theory \cite{HW,W2} one starts
from the strongly
coupled regime of $E_8\times E_8$ heterotic string theory and
 interprets it
as M-theory on an orbifold $\RR^{10}\times S^1/\ZZ_2$.
After compactification on a Calabi-Yau three-fold one arrives
at solutions which may be considered as a pair of parallel 3-branes
with opposite
tension, and location at the orbifold planes.

After 1995 it became clear
from investigations in Type I string theory
that due to compactified higher dimensions the string scale $M_s$ can be
 much smaller
than the 4-dimensional Planck scale $M_{Pl(4)}=1.22\times 10^{19}$ GeV and
that it is  bounded from below only experimentally by the scale
of electroweak
interaction
$M_{EW} \lesssim M_s \lesssim M_{Pl(4)}$
\cite{W2,Ly1,sub-mill1,sub-mill1a,Tye,Iban2}.
As suggested by Arkani-Hamed et al
\cite{sub-mill1,sub-mill1a,sub-mill2} it is even possible to lower the
fundamental Planck scale $M_{Pl(4+D^{\prime })}$ of
the $(4+D^{\prime })-$dimensional
theory  down to the SM electroweak scale $M_{Pl(4+D^{\prime })}
\sim M_{EW} \sim 1$ TeV.
This allows for a solution of
the hierarchy problem not relying on supersymmetry or technicolor.
In this approach gravity can
propagate in all multidimensional bulk space whereas ordinary
SM fields are localized on a 3-brane with thickness in the extra dimensions
of order the Fermi length $L_F$.
As a result, the 4-dimensional Planck scale
of the external space is connected with the electroweak scale by the
relation
\be{1.1}
M_{Pl(4)}^2 \sim V_{D^{\prime}} M_{EW}^{(2+D^{\prime})}\, ,
\ee
where $V_{D^{\prime}}$ is the volume of the internal spaces.
Thus, the scale of the internal space
compactification is of order
\be{1.2}
a \sim V_{D^{\prime}}^{1/D^{\prime}}
\sim 10^{\frac{32}{D^{\prime}}-17} \mbox{cm}\, .
\ee
In this model physically acceptable values correspond to $D^{\prime}\ge 2$,
e.g. for $D^{\prime} =2$ the internal space scale of compactification is
$a\sim 10^{-1} \mbox{cm}$.

The stabilization of  extra dimensions (geometrical moduli stabilization)
 in models with sub-millimetre internal spaces was
considered in Refs. \cite{sub-mill2,sub-mill3} where the dynamics
of the conformal
excitations of the internal spaces near minima of an effective potential
have been investigated.
Due to the product topology of the $(4+D^{\prime })-$dimensional bulk
spacetime constructed from Einstein spaces with scale (warp) factors
depending only on the coordinates of the external 4-dimensional component,
the conformal excitations have the form of massive scalar fields living
in the external space.
Within the framework of multidimensional cosmological
models (MCMs) we investigated such excitations
in  \cite{GZ(PRD),Beersh,GKZ,GZ(CQG)}
and called them gravitational excitons. Later, since the
submillimetre weak-scale compactification approach these geometrical moduli
excitations are known as radions \cite{sub-mill2,sub-mill3}.

Recently  Randall and Sundrum  \cite{RS} proposed an
interesting construction for the solution
of the hierarchy problem localizing low energy
SM matter as well as low energy gravity on a 3-brane in a slice of
anti-de Sitter space $AdS_5$.
Subsequently, it has been shown that such a localization
of low energy physics can be also
achieved at the intersection of a system of $(n+2)-$branes in $AdS_{4+n}$
\cite{ADDK} allowing for an interpretation of
our observable universe  e.g. as
a defect in a higher dimensional brane crystal \cite{K2}. But the RS
proposal and its generalizations are not considered in the present paper.

The main goal of our present comments consists in a clarification
of conditions
which ensure the stabilization of the internal spaces in multidimensional
models with a minimal coupled scalar field as a matter source
(section \ref{stabilization}).
A general method
for the solution of such problems in models with an arbitrary number
of internal spaces was proposed in Ref. \cite {GZ(PRD)}.
There it was shown that the problem of the internal space stabilization
can be solved most easily in the Einstein frame (although it is clear
that if  stabilization takes place in the Einstein frame it will also take
place in the Brans-Dicke frame and vice versa). Our investigations
(see also \cite{Beersh,GKZ,GZ(CQG)}) show
that inflation of the external space which was maintained for some models
in earlier Refs. (see e.g. \cite{Ezawa1,Ezawa2}) is destroyed by
a required stabilization of the internal spaces.

On the other hand there are also papers devoted to inflation where
stabilization of the internal spaces was supposed a priori (see e.g.
\cite{KO1,KO2,KL}).
We would like to stress here that it is  necessary
to be rather careful
in this case because stabilization can destroy inflation. For example,
the
appearance of a negative effective cosmological constant,
which in some
models is  a
 necessary
condition for
 the internal spaces stabilization,
can either
destroy inflation at all or make problematic its succesfull completion.
This situation occures
e.g. in the simple toy model which we consider in section \ref{inflation}
of the present paper. We use this model
in order to show exactly under which conditions  stabilization
takes place in multidimensional cosmological models with a
minimal coupled
scalar field and to discuss briefly a possibility for inflation in these
models.

In the present paper most of the
 calculations
 are performed in the electroweak
fundamental scale approach.
In the
conclusion section \ref{conclusion}
we compare the corresponding results with
those for the Planck fundamental
scale approach
 and show that the transition from one approach to the other results in a
rescaling of the effective
cosmological constant $\Lambda_{eff}$ as well as of
gravitational exciton
masses $m_i$.
The corresponding
 rescaling prefactors
which appear due to the
 transition  (see eq. \rf{c.3})
lead to a different
functional dependence of $\Lambda _{eff}$ and $m_i$
on the
 compactification sizes of the internal spaces in
 the two approaches.
As result,
in the
Planck fundamental scale approach the
values of $\Lambda_{eff}$ and $m_i$
can be
much smaller than in the electroweak approach.
Finally,
we discuss  some bounds on the
parameters of the model which follow from
observable cosmological data. These bounds strongly depend on
the details of the
behavior of the inflaton and gravitational exciton fields after inflation,
e.g. on the times of their reheating and decay.

\section{Stabilization of the internal spaces \label{stabilization}}
\setcounter{equation}{0}

\bigskip
We consider a cosmological model with metric
\begin{equation}
\label{2.1}g=g^{(0)}+\sum_{i=1}^ne^{2\beta ^i(x)}g^{(i)},
\end{equation}
which is defined on a manifold with product topology
\begin{equation}
\label{2.2}M=M_0\times M_1\times \dots \times M_n,
\end{equation}
where $x$ are some coordinates of
the $D_0 =(d_0+1)$ - dimensional manifold $M_0$ and
\begin{equation}
\label{2.3}g^{(0)}=g_{\mu \nu }^{(0)}(x)dx^\mu \otimes dx^\nu .
\end{equation}
Let manifolds $M_i$ be $d_i$ - dimensional Einstein spaces with metric $%
g^{(i)},$  i.e.
\begin{equation}
\label{2.4}R_{mn}\left[ g^{(i)}\right] =\lambda ^ig_{mn}^{(i)},\qquad
m,n=1,\ldots ,d_i
\end{equation}
and
\begin{equation}
\label{2.5}R\left[ g^{(i)}\right] =\lambda ^id_i\equiv R_i.
\end{equation}
In the case of constant curvature spaces parameters $\lambda ^i$ are
normalized as $\lambda ^i=k_i(d_i-1)$ with $k_i=\pm 1,0$. Later on we shall
not specify the structure of the spaces $M_i$. We require only $M_i$ to be
compact spaces with arbitrary sign of curvature.

With total dimension $D= D_0 +\sum_{i=1}^nd_i$,\, $\kappa_D ^2$ a
$D-$dimensional
gravitational constant, $\Lambda $ - a $D-$dimensional cosmological
constant and $S_{YGH}$ the standard York - Gibbons - Hawking boundary term
\cite{York,GH}, we consider an action of the form
\begin{equation}
\label{2.6}S=\frac 1{2\kappa_D ^2}\int\limits_Md^Dx\sqrt{|g|}\left\{
R[g]-2\Lambda \right\}
-\frac12 \int\limits_Md^Dx\sqrt{|g|}\left( g^{MN}\partial_M \Phi
\partial_N \Phi + 2 U(\Phi )\right) + S_{YGH}\, ,
\end{equation}
where the minimal coupled scalar field $\Phi$ with an arbitrary potential
$U(\Phi)$
depends on the external coordinates $x$ only. This field can be understood
as a zero mode of a bulk field. Such a scalar field can naturally
originate also in non-linear
$D-$dimensional theories \cite{GZ(non-lin)} where metric ansatz \rf{2.1}
ensures its dependence on $x$ only.

Let $\beta^i_0$ be the scale of compactification of the internal spaces
at the present time and
\be{2.7}
V_{D^{\prime }}
 \equiv V_I\times v_0 \equiv \prod_{i=1}^n\int\limits_{M_i}d^{d_i}y
\sqrt{|g^{(i)}|} \times \prod_{i=1}^n e^{d_i\beta^i_0}
\ee
the corresponding total volume of the internal spaces
($[V_{D^{\prime }}] = {\mbox cm}^{D^{\prime}}$,
 $[V_I]=1$,
 where $D^{\prime} = D-D_0$ is the
number of extra dimensions). Instead of $\beta^i$ it is convenient to
introduce a shifted quantity:
\be{2.8}
\tilde \beta^i = \beta^i - \beta^i_0\, .
\ee
Then, after dimensional reduction action \rf{2.6} reads
\ba{2.9}
S&=&\frac 1{2\kappa _0^2}\int\limits_{M_0}d^{D_0}x\sqrt{|g^{(0)}|}%
\prod_{i=1}^ne^{d_i\tilde \beta ^i}\left\{ R\left[ g^{(0)}\right]
-G_{ij}g^{(0)\mu\nu }\partial _\mu \tilde \beta ^i\,\partial _\nu
\tilde \beta ^j+\right. \nn \\
&+& \left. \sum_{i=1}^n \tilde R_i e^{-2\tilde \beta^i}-2\Lambda -
g^{(0)\mu \nu}\kappa^2_D \partial_{\mu} \Phi \partial_{\nu} \Phi
-2\kappa^2_D U(\Phi)\right\} \, ,
\ea
where $\tilde R_i := R_i e^{-2\beta^i_0}$, $G_{ij} = d_i\delta_{ij} -
d_i d_j \; (i,j = 1,\dots ,n)$ is the midisuperspace metric \cite{IMZ,RZ}
and
\be{2.10}
\kappa^2_0 := \frac{\kappa^2_D}{V_{D^{\prime }}}
\ee
is the $D_0-$dimensional (4-dimensional) gravitational constant.
If we take the electroweek scale $M_{EW}$ and the Planck scale $M_{Pl}$
as fundamental ones for $D-$dimensional and 4-dimensional space-times
respectively:
\ba{2.11}
\kappa^2_D &=& \frac{8\pi}{M^{2+ D^{\prime}}_{EW}}\, , \nn \\
&\phantom{-} &  \\
\kappa^2_0 &=& \frac{8\pi}{M^2_{Pl}}\, , \nn
\ea
then we reproduce  eqs. \rf{1.1} and \rf{1.2}.

Action \rf{2.9} describes a generalized $\sigma-$model
with target space metric $G_{ij}$ where the scale factors $\beta^i$ play
the role of scalar fields. The problem of the internal space stabilization
is reduced now to the investigation of the dynamics of these fields. Most
easily this can be done in the Einstein frame. For this purpose we perform
a conformal transformation
\begin{equation}
\label{2.12} g_{\mu \nu }^{(0)}= \Omega^2 \tilde g_{\mu \nu
}^{(0)} := {\left( \prod_{i=1}^ne^{d_i\tilde \beta ^i}\right) }
^{\frac{-2}{D_0-2}}
\tilde g_{\mu \nu }^{(0)}
\end{equation}
which yields  \cite{GZ(PRD)}
\begin{equation}
\label{2.13}S=\frac 1{2\kappa _0^2}\int\limits_{M_0}d^{D_0}x\sqrt{|\tilde
g^{(0)}|}\left\{ \tilde R\left[ \tilde g^{(0)}\right] -\bar G_{ij}
\tilde g^{(0)\mu \nu }\partial _\mu \tilde \beta ^i\,
\partial _\nu \tilde \beta ^j- \tilde g^{(0)\mu \nu}\kappa^2_D
\partial_{\mu} \Phi \partial_{\nu} \Phi - 2U_{eff}\right\} ,
\end{equation}
where $\bar G_{ij}=d_i\delta _{ij}+\frac 1{D_0-2}d_id_j$
and
\begin{equation}
\label{2.14}U_{eff}[\tilde \beta ,\Phi ] =
{\left( \prod_{i=1}^ne^{d_i\tilde \beta ^i}\right) }^{-\frac
2{D_0-2}}\left[ -\frac 12\sum_{i=1}^n\tilde R_ie^{-2\tilde \beta ^i}+\Lambda
+\kappa_D^2 U(\Phi ) \right]
\end{equation}
is the effective potential.

With the help of a regular coordinate transformation
$\varphi =Q\beta ,\quad \beta =Q^{-1}\varphi$
midisuperspace metric (target
space metric) $\bar G$ can be transformed to a pure Euclidean form:
$\bar G_{ij}d\beta ^i\otimes d\beta ^j=\sigma _{ij}d\varphi ^i\otimes
d\varphi ^j=\sum_{i=1}^nd\varphi ^i\otimes d\varphi ^i,\quad
\sigma ={\rm diag\ }(+1,+1,\ldots ,+1)$.
An appropriate transformation $Q:\
\beta ^i\mapsto \varphi ^j=Q_i^j\beta ^i$ can be found e.g. in
\cite{GZ(PRD)}.
We note that in the case of one internal space $(n=1)$
this transformation is reduced
to a simple redefinition
\begin{equation}
\label{2.15}\varphi \equiv \varphi^1 := \pm \sqrt{\frac{d_1(D-2)}{D_0-2}}
\tilde \beta ^1\
\end{equation}
which yields
\begin{equation}
\label{2.16}
U_{eff}[\varphi , \Phi ] = e^{2\varphi \sqrt{ \frac{d_1}{(D-2)(D_0-2)}}}
\left[ -\frac 12\tilde R_1e^{2\varphi \sqrt{ \frac{D_0-2}{d_1(D-2)}}}
+\Lambda +\kappa_D^2 U(\Phi)\right] \, .
\end{equation}
(For definiteness we use the minus sign in eq. \rf{2.15}.)

It is clear now that stabilization of the internal spaces can be achieved
iff the effective potential $U_{eff}$ has a minimum with respect to fields
$\tilde \beta^i$ (or fields $\varphi^i$). In general it is possible
for potential $U_{eff}$
to have more than one extremum.
But it can be easily seen that for the model
under consideration we can get one extremum only. Let us find conditions
which ensure a minimum at $\tilde \beta = 0$.

The extremum condition yields:
\be{2.17}
\left.\frac{\partial U_{eff}}{\partial \tilde \beta^k}\right|_{\tilde \beta
=0}
=0
\Longrightarrow
\tilde R_k = -\frac{d_k}{D_0-2}\left( \sum_{i=1}^n \tilde R_i -2(\Lambda
+ \kappa^2_D U(\Phi) )\right)\, .
\ee
The left-hand side of this equation is a constant
but the right-hand side is a
dynamical function. Thus, stabilization of the internal spaces in such type
of models is possible only when the effective potential has also a minimum
with
respect to the scalar field $\Phi$ (in Ref. \cite{BZ} it was proved that
for this model the only possible solutions with static internal spaces
correspond to the case when the
minimal coupled scalar field is in its
extremum position too).
Let $\Phi_0$ be the minimum position for field $\Phi$.
{}From the structure of the effective potential \rf{2.14} it is clear that
minimum positions of the potentials $U_{eff}[\tilde \beta ,\Phi]$ and
$U(\Phi)$ with respect to field $\Phi$ coincide with each other:
\be{2.18}
\left. \frac{\partial U_{eff}}{\partial \Phi}\right|_{\Phi_0} = 0
\Longleftrightarrow
\left. \frac{\partial U(\Phi )}{\partial \Phi }\right|_{\Phi_0} = 0\, .
\ee
Hence, we should look for parameters which ensure a minimum of $U_{eff}$
at the point ${\tilde \beta^i = 0, \Phi = \Phi_0}$. Eqs. \rf{2.17}
show that there exists a fine tuning condition for the scalar curvatures
of the internal spaces:
\be{2.19}
\frac{\tilde R_k}{d_k} = \frac{\tilde R_i}{d_i}\, ,
\quad (i,k = 1,\dots ,n)\, .
\ee
Introducing  the auxiliary quantity
\be{2.20}
\tilde \Lambda \equiv
\Lambda + \left. \kappa^2_D U(\Phi)\right|_{\Phi_0} \, ,
\ee
we get the useful relations
\be{2.21}
\Lambda_{eff} :=\left. U_{eff}\vphantom{\int} \right|_
{\tilde \beta^i =0,\atop \Phi = \Phi_0}\;
=\, \frac{D_0-2}{D-2}\tilde \Lambda\, =\,
\frac{D_0-2}{2}\frac{\tilde R_k}{d_k}\, ,
\ee
which show that $\sign \Lambda_{eff} =\sign \tilde \Lambda =\sign R_k$.
It is clear that $\Lambda_{eff}$ plays the role of an effective cosmological
constant in the external space-time. For the masses
of the normal mode excitations
of the internal spaces (gravitational excitons) and of the scalar field near
the extremum position we obtain respectively \cite{GZ(PRD)}:
\ba{2.22}
m_1^2 &=& \dots =m_n^2 = -\frac{4\Lambda_{eff}}{D_0-2}=-2\frac{\tilde R_k}
{d_k} > 0\, , \nn \\
&\phantom{-} & \\
m_{\Phi}^2 &:=& \left. \frac{\partial^2 U(\Phi )}{\partial \Phi^2}
\right|_{\Phi_0}\, .\nn
\ea
These equations show that for our specific model
a global minimum can only exist
in the case of compact internal spaces with negative curvature
$R_k <0\; (k=1,\dots ,n)$.
The effective cosmological constant is negative also: $\Lambda_{eff} <0$.
Obviously, in this model it is impossible
to trap the internal spaces at a minimum of $U_{eff}$ if they are tori 
($\tilde R_i = 0$)
because for Ricci-flat internal spaces the effective potential
has no minimum at all. 
Eqs. \rf{2.21} and \rf{2.22} show also that 
a stabilization by trapping
takes place only for $\tilde \Lambda <0$\footnote{An
interesting scenario for a dynamical stabilization of the
internal space  was proposed
in Ref. \cite{Maz} for a model with $\tilde \Lambda = \tilde R_i = 0$.
If, at some stage of the Universe evolution, the inflaton field $\Phi$
reached its zero minimum and was frozen out, then there exists a solution
$\tilde \beta \longrightarrow 0$ for times $t \longrightarrow \infty$ which
corresponds to a dynamical stabilization of the internal space.
 However,
the inflaton field is never frozen out completely and its dynamics can
destabilize the internal space. An investigation of this problem in
collaboration with Anupam Mazumdar will be presented soon in a common
paper.}.
This means that the minimum of the scalar field potential should be
negative $U(\Phi_0)<0$ for non-negative bare cosmological
constant $\Lambda \ge 0$ or it should satisfy inequality
$\kappa^2_D U(\Phi_0)<|\Lambda |$ for $\Lambda <0$.

For small fluctuations of the normal modes in the vicinity
of the minima of the effective potential action \rf{2.13} reads
\begin{eqnarray}\label{2.23}
S & = & \frac{1}{2\kappa _0^2}\int \limits_{M_0}d^{D_0}x \sqrt
{|\tilde g^{(0)}|}\left\{\tilde R\left[\tilde g^{(0)}\right] - 2\Lambda
_{eff}\right\} - \\
\ & - & \frac{1}{2}\int \limits_{M_0}d^{D_0}x \sqrt
{|\tilde g^{(0)}|}\left\{\sum_{i=1}^n \left( \tilde g^{(0)\mu \nu}
\psi ^i_{,\mu}\psi^i_{,\nu} + m_i^2\psi ^i\psi ^i\right)
+ \tilde g^{(0)\mu \nu} \phi_{,\mu} \phi_{,\nu} +
m_{\phi}^2 \phi \phi \right\}\, . \nn
\end{eqnarray}
(For convenience we use here the normalizations:
$\kappa^{-1}_0 \tilde \beta \rightarrow \tilde \beta$ and
$\sqrt{V_{D^{\prime }}}(\Phi -\Phi_0) \rightarrow \phi$.)
Thus, conformal excitations of the metric of the internal spaces behave as
massive scalar fields developing on the background of the external
spacetime. In analogy with excitons in solid state physics where they are
excitations of the electronic subsystem of a crystal, we called
the excitations of the
subsystem of internal spaces  gravitational excitons
\cite{GZ(PRD)}. Later, since \cite{sub-mill2,sub-mill3} these
particles are also known as radions.

\section{Inflation of the external space \label{inflation}}
\setcounter{equation}{0}
In this section we discuss briefly the possibility for inflation
in the external
space of our model. We perform the analysis
in the Einstein frame where the effective
theory is described by action \rf{2.13} and inflation depends on
the form of potential \rf{2.14}.

For simplicity we consider a model with only
one internal space and an
effective potential given by equation \rf{2.16}. All our conclusions
can be easily generalized to a model with $n$ internal spaces.

First, we consider region
\be{3.1}
e^{2\varphi \sqrt{\frac{D_0-2}{d_1(D-2)}}}\gg |\Lambda
+\kappa^2_D U(\Phi)|\, ,
\ee
where the effective potential reads
\be{3.2}
U_{eff} \approx \frac12 |\tilde R_1|e^{2\varphi \sqrt{\frac{D-2}
{d_1(D_0-2)}}}\, .
\ee
It is well known \cite{LL} that for models with potential $U(\varphi) \sim
Ae^{\lambda \varphi}$ the scale factor behaves as $\tilde a \sim \tilde
t^{2/\lambda^2}$ and power law inflation takes place if $\lambda^2 <2$.
In our case we have
\be{3.3}
\lambda^2 = \left. \frac {4(D-2)}{d_1(D_0-2)}\right|_{D_0=4}
= 2\left( 1+\frac{2}{d_1}
\right) >2\,
\ee
and power law inflation is impossible in this region of the model.
For the model with $n$ internal spaces
the assisted inflation proposed in Ref. \cite{LMS} is impossible also
in this region because of the form of
the effective potential (it is impossible to split the effective potential
into a sum of $n$ terms where each of them depends on one scalar field
only).

Second, we consider the region near the minimum of the effective potential.
In the scenario of
assisted chaotic inflation \cite{KO1,KO2,KL} with a sufficiently large
number of scalar fields $\psi^i$  inflation occurs at scales much less
than Planck scale: $|\psi^i| \ll 1$. In our model the effective action
for these fields is given by eq. \rf{2.23} and
it has the typical form of an action
allowing for this type of inflation. Therefore, it is of interest to
investigate the possibility for assisted chaotic inflation here.
Unfortunately,
for our particular model the internal space stabilization takes place
only for negative effective cosmological constant. This destroys
inflation because, as it follows from eq. \rf{2.22} $m^2_i \sim
|\Lambda_{eff}|$,  the energy density of the potential $U_{eff}$ is not
sufficient for inflation. There is also another drawback of theories with
negative cosmological constant. Even if they have a period of inflation
there is a problem of succesful completion of it. We shall return to this
problem in the next section.

Third, we consider the region
\be{3.4}
\kappa^2_D U(\Phi ) \gg \Lambda + \frac12 |\tilde R_1 |
e^{2\varphi \sqrt{\frac{D_0-2}{d_1(D-2)}}}\, ,
\ee
where the effective potential reads
\be{3.5}
U_{eff} \approx e^{2p\varphi}\kappa^2_D U(\Phi )\, ,
\quad p := \sqrt{\frac{d_1}{(D-2)(D_0-2)}}\, .
\ee
For models with $n+1$ scalar fields the slow roll conditions
are \cite{GZ(non-lin)}:
\be{3.6}
\epsilon \approx \frac{1}{2U^2_{eff}}\sum_{i=1}^{n+1}
\left(\frac{\partial U_{eff}}{\partial \varphi^i}\right)^2
\ee
and
\be{3.7}
\eta _i \approx - \epsilon + \frac{1}{U_{eff}}
\sum_{j=1}^{n+1}
\frac{\partial^2 U_{eff}}{\partial \varphi^i \partial \varphi^j}
\left(\left.\frac{\partial U_{eff}}{\partial \varphi^j}\right/
\frac{\partial U_{eff}}{\partial \varphi^i}\right)  \; , \quad
i=1,\dots ,n+1\, .
\ee
Inflation is possible if these parameters are small: $\epsilon , |\eta_i |
< 1$. For potential \rf{3.5} we get:
\ba{3.8}
\epsilon &\approx & \eta_1 \approx 2p^2 + \epsilon_{\Phi}\, ,\nn \\
\eta_2 &\approx & 2p^2 + \eta_{\Phi}\, ,
\ea
where $\epsilon_{\Phi} := \frac12 \left( \frac{U^{\prime}(\Phi )}
{U(\Phi )}\right)^2$ and $\eta_{\Phi } := -\epsilon_{\Phi } +
\frac{U^{\prime \prime }(\Phi )}{U(\Phi )}$. Because of
\be{3.9}
\left. 2p^2\right|_{D_0=4} =  1-\frac{2}{d_1+2} <1\, ,
\ee
 inflation  is possible in this region  if
\be{3.10}
\epsilon_{\Phi },\, \eta_{\Phi} \ll 1\, .
\ee
Thus, the scalar field $\Phi$ can act as inflaton and
drive the inflation of the external
space if its potential in region \rf{3.4} satisfies conditions \rf{3.10}.
It is clear that estimates \rf{3.9} and \rf{3.10} are rather crude
and they show only the principal possibility for inflation to occur.
For each particular form of $U(\Phi )$ a detailed analysis of the dynamical
behavior of the fields in this region should be performed to confirm
inflation.
Obviously, if the inflation in our model is realized it takes place
before the stabilization of the internal spaces.
In the case of constant scalar
field $\Phi = \const $ or its absence
 inflation of the external space in our model is impossible at all.

\section{Discussion and conclusions \label{conclusion}}
\setcounter{equation}{0}

\bigskip
In the present paper we considered the possibility for
 stabilization of the internal space
 and inflation in the external space using as example
 a multidimensional
cosmological toy model with minimal coupled scalar field as matter source.
The calculations above were
 performed in a
model with  the electroweek scale $M_{EW}$
as  fundamental
scale of the $D-$dimensional
theory (see eq. \rf{2.11}).
Clearly, it is also  possible  to choose the Planck scale
as the fundamental scale.

For this purpose  we will not fix the compactification scale
 of the internal
spaces at the present time. We consider them as free parameters of
the model and demand only that $L_{Pl} < a_{(0)i} = e^{\beta^i_0} <
L_{F}$. So, we shall not transform $\beta^i$ to $\tilde \beta^i$.
In this case, after dimensional reduction of action \rf{2.6}
the effective $D_0-$dimensional
gravitational constant $\kappa^2_0$ is defined as
\be{c.1}
\frac{1}{\kappa^2_0} = V_I\frac{\left(L_{Pl}\right)^{D^{\prime}}}
{\kappa^2_D}\, .
\ee
At the other hand there holds
 $\kappa^2_0= 8\pi /M_{Pl}^2$ (for $D_0=4$).
Thus, $\kappa^2_D = 8\pi V_I/M^{(2+D^{\prime})}_{Pl}$, so that
the Planck scale
becomes the fundamental scale of $D-$dimensional
theory. In this approach eqs. \rf{2.9}, \rf{2.12} - \rf{2.16}
preserve their form with only substitutions $\tilde \beta \longrightarrow
\beta$ and $\tilde R_i \longrightarrow R_i$.

The analysis of the internal space stabilization  shows that the fine tuning
condition \rf{2.19} is not changed:
\be{c.2}
\frac{R_k}{d_k} e^{-2\beta^k_0}= \frac{R_i}{d_i} e^{-2\beta^i_0}\, ,
\quad i,k=1,\dots ,n
\ee
and the masses squared of the gravitational excitons and the effective
cosmological constant are shifted by the same prefactor:
\ba{c.3}
m^2_i &\longrightarrow &\left(\prod_{i=1}^n e^{d_i\beta_0^i}\right)^
\frac{-2}{D_0-2} m_i^2
= -2\left(\prod_{i=1}^n e^{d_i\beta_0^i}\right)^
\frac{-2}{D_0-2} \frac{R_i}{d_i}
\, , \nn \\
&\phantom{-}& \\
\Lambda_{eff} &\longrightarrow &\left(\prod_{i=1}^ne^{d_i\beta_0^i}\right)^
\frac{-2}{D_0-2} \Lambda_{eff}
=\frac{D_0-2}{2}
\left(\prod_{i=1}^n e^{d_i\beta_0^i}\right)^
\frac{-2}{D_0-2}\frac{R_i}{d_i}
\, . \nn
\ea
 For example, in the one-internal-space case we get
the estimate\footnote{We use  standard Planck length unit conventions with
$[m]=\mbox{\rm cm}^{-1}$ and the corresponding shorthand, \\
e.g. $ m_1^2\sim (a_{(0)1})^{-(D-2)}\equiv
\left(\frac{a_{(0)1}}{L_{Pl}}\right)^{-(D-2)}L_{Pl}^{-2}.$ }
 \cite{GZ(PRD)}:
\be{c.4}
|\Lambda_{eff}| \sim m_1^2 \sim
\left. \left(e^{-\beta^1_0}\right)^{2\frac{D-2}{D_0-2}}\right|_{D_0=4}
= \left(a_{(0)1}\right)^{-(D-2)}\, .
\ee
This expression shows that due to the power $(2-D)$
the effective cosmological constant and the masses of the gravitational
excitons can be very far from the planckian values even for scales of
compactification of the internal spaces close to the Planck length.

Another important note consists in the observation that the Einstein frame
metrics of the external spacetime in both approaches are equivalent to each
other up to a numerical prefactor:
\be{c.4a}
\left. \tilde g^{(0)}_{\mu \nu}\right|_{EW}
=v_0^{-2/(D_0-2)}\left.\tilde g^{(0)}_{\mu \nu}\right|_{Pl}\, .
\ee
 Equation
\rf{2.12} shows that in the electroweak approach the Brans-Dicke and
Einstein
scales coincide with each other at the point of stabilization: $\tilde
\beta^i = 0 \Longrightarrow \Omega = 1$. In the Planck
fundamental scale approach this has
place when the internal scale factors are equal to the Planck length:
$\beta^i =0 \Longrightarrow \Omega =1$.
This does not mean that in the latter approach
the  stabilization of the internal space takes place at the Planck length.
Depending on the concrete  form of
the effective potential $U_{eff}$ its minimum position/stabilization
point $\beta ^i_0$
can  be located at much larger scales
$L_{Pl}\ll a_{(0)i}=e^{\beta ^i_0}L_{Pl}$.
Generally speaking, we should not exclude also
a possibility for internal spaces to change very slowly with time. In this
case $\beta^i_0 $ is not so strictly defined as for models with the internal
space stabilization in minima of the effective potential.

Let us return to the comparision of the electroweak and the Planck scale
approaches. From eqs. \rf{c.3} it is clear that the
reason for the rescaling/lightening of the effective cosmological constant
as well as of the gravitational exciton masses in the Planck scale approach
consists in the prefactor
$\left(\prod_{i=1}^n e^{d_i\beta_0^i}\right)^\frac{-2}{D_0-2}$.
 In spite of the smallness of the internal space sizes
in the Planck fundamental scale approach
($L_{Pl} < a_{(0)1} < L_F$)
in comparison with the sizes in
the electroweak fundamental scale approach
($a_{(0)1}\sim 10^{-1}$cm for $D^{\prime }=2$
and $a_{(0)1}\to 10^{-17}$cm for $D^{\prime }\to \infty $),
the prefactors in eqs. \rf{c.3}
can considerably reduce the values of $\Lambda _{eff}$ and $m_i $
making them much smaller then in the electroweak approach.

Let us compare now some estimates following from the electroweak as well as
from the Planck fundamental scale approaches. (We use the obvious subscripts
$EW$ and $Pl$ respectively.)

In the first case, the scale of the internal space compactification is given
by formula \rf{1.2}. We take for definiteness the total number of dimensions
$D=6$ and $D=10$ and obtain respectively the following scales of
compactification:
$a_{(0)1}\sim 10^{-1}{\mbox cm}$ for $D=6$ and
$a_{(0)1}\sim 10^{-9}{\mbox cm}$ for $D=10$. Then,
from eqs. \rf{2.21} and \rf{2.22} we get:
\be{c.5}
\left. |\Lambda_{eff} |\vphantom{\int} \right|_{EW}
\sim \frac{1}{a^2_{(0)1}} \sim
\left\{\begin{array}{rcl}
10^2{\mbox cm}^{-2}&\sim & 10^{-64}\Lambda_{Pl}\, , \quad D=6\\
10^{18}{\mbox cm}^{-2}&\sim & 10^{-48}\Lambda_{Pl}\, , \quad D=10\\
\end{array}\right.
\ee
and
\be{c.6}
\left. m_1\vphantom{\int} \right|_{EW}
\sim \frac{1}{a_{(0)1}} \sim
\left\{\begin{array}{rcl}
10^{-32}M_{Pl}&\sim & 10^{-4}eV\, , \quad D=6\\
10^{-24}M_{Pl}&\sim & 10^4eV\, , \quad D=10\\
\end{array}\right. \, .
\ee

In the second case, the scale of compactification is not fixed, but a
free parameter. We demand only that it should be smaller then the Fermi
length. For definiteness let us use $a_{(0)1} \sim 10^{-18}{\mbox cm}$.
Then, from eq. \rf{c.4}
we obtain:
\be{c.7}
\left. |\Lambda_{eff} |\vphantom{\int} \right|_{Pl}
\sim a^{-(D-2)}_{(0)1} \sim
\left\{\begin{array}{rcl}
10^6{\mbox cm}^{-2}&\sim & 10^{-60}\Lambda_{Pl}\, , \quad D=6\\
10^{-54}{\mbox cm}^{-2}&\sim & 10^{-120}\Lambda_{Pl}\, , \quad D=10\\
\end{array}\right.
\ee
and
\be{c.8}
\left. m_1\vphantom{\int} \right|_{Pl}
\sim a^{-(D-2)/2}_{(0)1} \sim
\left\{\begin{array}{rcl}
10^{-30}M_{Pl}&\sim & 10^{-2}eV\, , \quad D=6\\
10^{-60}M_{Pl}&\sim & 10^{-32}eV\, , \quad D=10\\
\end{array}\right.
\ee

Estimates \rf{c.5} and \rf{c.7} show that for the electroweak scale
the effective cosmological constant is much greater than the present
day observable limit
$\Lambda \le 10^{-122}\Lambda_{Pl} \sim 10^{-57}{\mbox cm}^{-2}$
(for our model
$\left. |\Lambda_{eff}|\right|_{EW}\ge 10^2 \, \mbox{\rm cm}^{-2}$),
whereas in the
Planck scale approach we can satisfy this limit
even for very small compactification scales.
For example, if we demand in accordance with observations $|\Lambda_{eff}|
\sim 10^{-122}\Lambda_{Pl}$ then eq. \rf{c.4} gives a
compactification  scale  $a_{(0)1} \sim 10^{122/(D-2)}L_{PL}$.
Thus, $a_{(0)1} \sim 10^{15}L_{Pl} \sim 10^{-18}{\mbox cm}$ for $D=10$ and
$a_{(0)1} \sim 10^{5}L_{Pl} \sim 10^{-28}{\mbox cm}$ for $D=26$, which
does not contradict to observations because for this approach the
scales of compactification should be
$a_{(0)1} \le 10^{-17}{\mbox cm}$. Assuming an estimate
$\Lambda_{eff}\sim 10^{-122}L_{Pl}$,
we automatically get from eq. \rf{c.4} the value
of the gravitational exciton mass: $m_1 \sim 10^{-61}M_{Pl} \sim
10^{-33}eV \sim 10^{-66}{\mbox g}$ which is extremely light. Nevertheless
such light particles are not in contradiction with the observable Universe,
as we shall
show below.

Similar to the Polonyi fields in spontaneously broken supergravity
\cite{CFKRR,ENQ} or moduli fields in the hidden sector of SUSY
\cite{BBSMS,CCQR,BKN,BBS} the gravitational excitons are WIMPs
(Weakly-Interacting Massive Particles \cite{KT})
because their coupling to the
observable matter is suppresed by powers of the Planck scale. In Ref.
\cite{GSZ} we show that the decay rate of the gravitational excitons with
mass $m_{\varphi}$ is $\Gamma \sim m_{\varphi}^3/M_{Pl}^2$ as for Polonyi
and moduli fields.

Let us assume for a moment that after inflation the inflaton field $\phi$
has already decayed and produced the main reheating
 of the Universe.
For our model it may happen if $m_{\phi} \gg m_{\varphi}$ and the inflaton
field starts to
oscillate and decay much earlier than the $\varphi-$field
(coherent oscillations of field $\phi$ with mass $m_{\phi}$ usually start
when the Hubble constant $H \le m_{\phi}$). The Universe is radiation
dominated in this period and the Hubble constant is defined by $H \sim
T^2/M_{Pl}$. After the temperature is fallen to the value $T_{in} \sim
\sqrt{m_{\varphi}M_{Pl}}$ the scalar field\footnote
{Here, $\varphi = \pm \frac{M_{Pl}}{\sqrt{8\pi}}\sqrt{\frac{d_1(D-2)}
{D_0-2}}\beta^1$ where $\beta^1$ is the logarithm of the internal space
scale factor: $a_1=e^{\beta^1}L_{Pl}$. If stabilization  occurs at
$a_{(0)1} \sim 10^{-n}L_{Pl}, \; (0<n<18)$, then it corresponds to the
minimum position $\varphi_0 = \mp \frac{n \ln 10}{\sqrt{8\pi }}
\sqrt{\frac{d_1(D-2)}{D_0-2}}M_{Pl}$.}
$\varphi$ begins to oscillate coherently around the minumim and its density
evolves as $T^3$ \cite{ENQ,PWW}:
\be{c.9}
\rho_{\varphi}(T) = \rho_{\varphi}(T_{in})\left(T/T_{in}\right)^3
= m^2_{\varphi}\varphi^2_{in}(T_{in})\left(T/T_{in}\right)^3\, ,
\ee
where $\varphi_{in} := (\varphi - \varphi_{0})_{in}$ is the amplitude of
initial oscillations of the field $\varphi$ near the minimum position.
It is clear that for the extremely light particles we can neglect their
decay
($\Gamma_{\varphi} \approx 0$).
Then, because the ratio $\rho_{\varphi}/ \rho_{rad}$  increases as
$1/T$, at some themperature the Universe will be dominated (up to present
time) by the energy density of the coherent oscillations. We can easily
estimate the mass of the gravitational excitons which overclose the
Universe.
Assuming that at present
time $\rho_{\varphi} \lesssim \rho_c$, where $\rho_c$
is the critical density of
the present day Universe, we obtain\footnote{See also Note added.}
\be{c.10}
m_{\varphi} \lesssim 10^{-56} M_{Pl} \left(\frac{M_{Pl}}{\varphi_{in}}
\right)^4\, .
\ee
Usually, it is assumed
that $\varphi_{in} \sim O (M_{Pl})$ although it depends
on the form of $U_{eff}$ and can be considerably less than $M_{Pl}$. If we
put $\varphi_{in} \sim O(M_{Pl})$ then excitons with masses $m_{\varphi}
\lesssim 10^{-28}{\mbox eV}$ will not overclose the Universe \cite{ENQ,BKN}.
If $\varphi_{in}
\ll M_{Pl}$ this estimate will be not so severe. We see that our
mass $m_{\varphi} \sim 10^{-33}{\mbox eV}$ satisfies
the most severe estimate.
It can be considered as  hot dark matter which negligibly contributes to the
total amount of  dark matter and does not contradict to the model of
cold dark matter.

Of course, as it follows from eqs. \rf{c.4} and \rf{c.8} the mass
$m_{\varphi}$ could be considerably  heavier than $10^{-33}{\mbox eV}$ but
as result we would arrive at an effective cosmological constant greater
than the
observable one (see eqs. \rf{c.7} and \rf{c.8} for $D=6$ and $a_{(0)1} \sim
10^{-18}{\mbox cm}$) and we would need a mechanism for
its reduction to the observable
value. An example for such a reduction of the cosmological constant
was proposed in
\cite{BBS} for SUSY breaking models with moduli masses $m \sim 10^{-2} -
10^{-3}{\mbox eV}$. Such masses we get also in our model if we take for the
Planck scale approach $D=6$ and $a_{(0)1} \sim 10^{-18}{\mbox cm}$ (see
\rf{c.8}). For these particles we cannot neglect the decay rate
$\Gamma_{\varphi}$ which results in converting of the coherent oscillations
into radiation. In this case the Universe has a further reheating to the
themperature \cite{ENQ,CCQR}
\be{c.11}
T_{RH} \sim \sqrt{\frac{m^3_{\varphi}}{M_{Pl}}}\, .
\ee
For $m_{\varphi} \sim 10^{-2}$ eV the reheating temperature
$T_{RH} \sim 10^{-23}\mbox{ MeV} \ll 1\mbox{ MeV}$  is much less than
the temperature $T \sim 1{\mbox MeV}$ at which the nucleosynthesis begins.
Thus, either decaying particles should have
masses $m_{\varphi }>10^4{\mbox GeV}$ to get
$T_{RH} > 1{\mbox MeV}$ or we should get rid off such particles before
nucleosynthesis. The latter can be achieved if the decay rate becomes
larger.
In \cite{BBS} it was proposed that at a very early stage of the Universe
evolution (after inflation) WIMPs collapse into stars (e.g. modular stars)
where their field strength could be very large and leads to a substantial
enhancement  of the decay into ordinary particles.
For example, in Ref. \cite{GSZ} it is shown that gravexcitons have
a coupling to photons of the form $\frac{\varphi}{M_{Pl}}F^2_{\mu \nu}$.
In the core of such stars the gravexciton amplitude $\varphi$ might be much
larger than $M_{Pl}$, enhancing the coupling of this field to photons and
leading to explosions of these stars into  bursts of photons.

As it follows from eqs. \rf{2.22} and \rf{c.6}, in the electroweak approach
gravexciton masses should satisfy the inequality
$m_{\varphi } \gtrsim 10^{-4}{\mbox eV}$.
If the above mentioned  mechanism of the gravexciton
energy dilution due to
modular star explosions or due to some other reasons\footnote{For example,
in \cite{sub-mill3} for this purpose  a short period of
late inflation was proposed which should be followed by a reheating.
However, it is necessary to be rather
careful to avoid the generation of quantum fluctuations of gravexcitons
during
inflation again \cite{GLV}.} does not work,
the bound $m_{\varphi } \gtrsim 10^4{\mbox GeV}$
is valid and leads to the large $D^{\prime}$ limit ($D^{\prime} \gg 30$)
with a
scale of compactification $a_{(0)1} \sim \sqrt{D^{\prime}}m_{\varphi}^{-1}$
(see
\rf{2.22}). Thus for $D^{\prime} \sim 100$ and $m_{\varphi } \sim
10^4{\mbox GeV}$ we
get $a_{(0)1} \sim 10^{-17}$ cm which is not in strong contradiction
with the value $a_{(0)1} \sim 10^{-16,7}{\mbox cm}$ which follows
from eq. \rf{1.2}.
It is clear that in this approach an increasing of the mass by one order
requires an increasing of the number of  internal dimensions by two orders.

Above, we considered the case  $m_{\phi} \gg m_{\varphi}$ when the inflaton
field starts to oscillate coherently much earlier than the scale factors
of the
internal spaces. Let us suppose now that $m_{\phi} \sim m_{\varphi} \equiv
m$.
Thus, the inflaton $\phi$ and gravexciton $\varphi$ fields start to
oscillate
coherently at the same time $t_{in}$ with approximately the same initial
amplitude $\phi_{in} \sim \varphi_{in}$.
by
At this time the Universe becomes
 matter dominated with $\rho_{\varphi} \sim \rho_{\phi } \sim 1/\tilde a^3$
where $\tilde a$ is the scale factor of the external space-time.
We assume also that the inflaton $\phi$ is not a  WIMP  and its
decay rate $\Gamma_{\phi} \sim \alpha^2_{\phi}m \gg \Gamma_{\varphi} \sim
m^3/M_{Pl}^2$. Thus the effective coupling $\alpha_{\phi}$ of the inflaton
field $\phi$ satisfies: $\alpha_{\phi} \gg m/M_{Pl}$. Because $m \ll
M_{Pl}$ the effective coupling $\alpha_{\phi}$ still may be much less than
1.

First, we consider the case when the gravexciton decay rate is negligibly
small: $\Gamma_{\varphi} \approx 0$.
Let $t_{RH}$ be the time of reheating due to inflaton decay and let us
suppose that all the inflaton
energy is converted into radiation ($\rho_{\phi}(t_{RH}) \sim \left.
\rho_{rad}\right|_{RH} \sim T^4_{RH}$). It can be easily seen that for
$t>t_{RH}$ the relative contribution of $\varphi$ to the energy density
starts
to increase as
\be{c.12}
\frac{\rho_{\varphi}(T)}{\rho_{rad}(T)} = \frac{T_{RH}}{T}\, .
\ee
Here, in the sudden decay approximation the reheating temperature, $T_{RH}$,
is defined by equating the Hubble constant with the rate of decay:
$H(t_D) \sim \Gamma_{\phi} \sim \alpha^2_{\phi} m$, where $t_D \sim t_{RH}$
is the decay time. Because $H^2(t_D) \sim M^{-2}_{Pl}\left.\rho_{\phi}
\right|_{t_D} \sim T^4_{RH}/M^2_{Pl}$ we get
\be{c.13}
m \sim \frac{1}{\alpha^2_{\phi}}\frac{T^2_{RH}}{M_{Pl}}\, .
\ee
This formula shows that to get the temperature
$T_{RH}>1{\mbox MeV}$, which is necessary for the nucleosynthesis,
the mass should satisfy the  inequality
\be{c.14}
m \gtrsim \frac{1}{\alpha^2_{\phi}} 10^{-16}{\mbox eV}\, .
\ee

{}At the other hand, at present time (which we denote by a subscript 0)
the condition that gravexcitons do not overclose the Universe reads:
$\left.\rho_{\phi}\right|_0
=\left.\left(T_{RH}/T_0\right)\rho_{rad}\right|_0
\lesssim \rho_c$ and gives a  second limit for the mass:
\be{c.15}
m \lesssim \frac{1}{\alpha^2_{\phi}}
\left( \frac{\rho_c}{\left.\rho_{rad}\right| _0}
\right)^2 \frac{T^2_0}{M_{Pl}}\, .
\ee
Inserting into this formula the present day values for the temperature $T_0$
and the critical energy density $\rho_c$ we obtain
\be{c.16}
m \lesssim \frac{1}{\alpha^2_{\phi}} 10^{-26}{\mbox eV}\, ,
\ee
which obviously is in contradiction to the previous estimate \rf{c.14}.

Second, to solve this problem we consider the possibility of a further
reheating due to  gravexciton decay: $\Gamma_{\varphi} \ne 0$.
In order to estimate the temperature at which this  decay occurs
we should take into account
that after the first reheating (with the temperature defined by \rf{c.13})
the Universe is  matter dominated because $\rho_{\varphi }/\rho_{rad}
= T_{RH}/T >1$ for $T < T_{RH}$ and for the Hubble constant holds
$H^2 \sim
\rho_{\varphi}/M^2_{Pl}$. Thus, equating the Hubble constant with the
 decay rate: $H (T^{\prime }_D) \sim \Gamma_{\varphi} \sim
m^3/M^2_{Pl}$ we obtain the temperature of the gravexciton
decay:
\be{c.17}
\left(T^{\prime }_D\right)^3 \sim \frac{1}{\alpha_{\phi}} \frac{m^{11/2}}
{M_{Pl}^{5/2}}\, .
\ee
In this scenario the temperatures of the gravexciton decay and the
reheating are
denoted by a prime to distinguish  them from the corresponding
temperatures of
inflaton decay and reheating.
In the sudden decay approximation the temperature of the second reheating
is obtained by equating the squared decay rate and
 the radiation energy
density just after reheating
(because $H^2(T^{\prime }_D) \sim \Gamma^2_{\varphi}
\sim \rho_{rad}/M^2_{Pl}$) which obviously leads again to eq. \rf{c.11}
(where $T_{RH}$ should be replaced by $T^{\prime }_{RH}$).
Again, for a successful
nucleosynthesis with $T^{\prime }_{RH} > 1{\mbox MeV}$ the mass should be
$m \gtrsim 10^4{\mbox GeV}$. The reheating from $T^{\prime }_D$
to $T^{\prime }_{RH}$
produces an entropy increase given by
\be{c.18}
\Delta = \left(\frac{T^{\prime }_{RH}}{T^{\prime }_D}\right)^3
\sim \alpha_{\phi}
\frac{M_{Pl}}{m} \gg 1\, ,
\ee
which is much greater than 1 because $\alpha_{\phi} \gg m/M_{Pl}$.
However, it may be much less (not so severe) than the
usual estimate \cite{ENQ}:
$\Delta \sim M_{Pl}/m$ because $\alpha_{\phi}$ may be much less than 1.
If we require $\Delta \lesssim 10^5$, as a maximal permissible factor for
the dilution of the high-temperature baryogenesis, we obtain the bound
$m \gtrsim \alpha_{\phi} 10^{-5} M_{Pl}$ and for $\alpha _{\phi } \ll 1$
this bound is not so strong as the usual one: $m \gtrsim 10^{14}$GeV.

Summarizing the discussion we see that in models,
where the coherent oscillation of gravitational excitons starts
in the radiation dominated era, the gravexcitons should be either extremely
 light (see eq. \rf{c.10})
or very heavy particles ($m_{\varphi }>10^4$GeV
for a successful nucleosynthesis;  in case that
the hot baryogenesis is taken into account: $m_{\varphi }>10^{14}$GeV).
In models, where  inflaton and  gravitational exciton start
their coherent oscillation at the same time, extremely light excitons
are forbidden. Heavier excitons
with masses $m_{\varphi }\gtrsim \alpha _{\varphi }10^{-5}M_{Pl}$ are
allowed
(for a successful nucleosynthesis and high-temperature baryogenesis).

As conclusion  we would like to note
  that in our toy model the stabilization of
the internal spaces is realized only when the effective cosmological
constant is negative (for both fundamental scale  approaches).
It is well known that for such models
inflation is never succeffully completed
\cite{BBS}, because in this case our (external) space
 has a turning point at its  maximal scale factor
 where it stops to expand and begins to contract. If the spatial curvature
of our Universe is non-negative (according to the
latest observational data it is zero),
then the internal scale factors cannot be freezed because solutions
$\beta^i = \beta^i_0 = \const$ correspond to a negative squared Hubble
constant: $H^2 = \frac13\left[ \frac12 \bar G_{ij}\dot \beta^i \dot \beta^j
+ U_{eff}\right] \Longrightarrow \frac13 \Lambda_{eff} < 0$. To describe the
post-inflationary stage for such models we should extend our consideration
including e.g.
additional perfect fluid terms into the action functional
which correspond to usual matter in the universe (see \cite{GZ(CQG)}
for the details of this method). Another  possible
generalization consists in an inclusion of additional terms which result in
a positive effective cosmological constant in accordance with recent
observational data \cite{PTW}. This can be achieved  e.g. with the help of
antisymmetric form-fields \cite{GZ(nl2)}. For these models the gravexciton
masses and the (positive) effective cosmological constant are defined by
equations similar to \rf{c.4}. Such models can solve the following three
important problems simultaneously:
they yield  stabilization of the internal spaces, allow for
 inflation of the external space, and lead to a positive observable
effective cosmological constant. In these models the mechanism of
lightening of the effective cosmological
constant as well as the gravitational exciton masses  will work also in
the Planck fundamental scale approach because eqs. \rf{c.3} are general
for this type of models.\\[1ex]


{\bf Note added}

Alexander Sakharov informed us about another upper bound on $m_{\varphi}$
following from isocurvature gravexciton fluctuations if $m_{\phi} \gg
m_{\varphi}$ because in this case gravexcitons on the stage of
inflation can be considered as massless particles. These isocurvature
fluctuations result in a CMBR anisotropy $\delta T/T$. The amplitude of
these fluctuations can be estimated as
$\delta \varphi \approx H_{inf}/2\pi$ and is connected
with $\delta T/T$ as follows: $\delta T/T \approx (\rho_{\varphi}/
\rho_c)(\delta \varphi/\varphi_{in}) \approx (\rho_{\varphi}/\rho_c)
(H_{inf}/2\pi \varphi_{in})$, where $H_{inf}$ is the Hubble constant
at the inflation stage. According to COBE data, $\delta T/T \lesssim
10^{-5}$ and $H_{inf} \approx 10^{-5}M_{Pl}$. Thus, we get following
limitation  on the gravexciton energy density at present time: $\rho_{\varphi}
\lesssim 2\pi \rho_c \varphi_{in}/M_{Pl}$. Substitution of eq. \rf{c.9}
into this limitation gives
\be{c.19}
m_{\varphi} \lesssim 10^{-55} M_{Pl} \left(\frac{M_{Pl}}{\varphi_{in}}
\right)^2\, .
\ee
So, if $\varphi_{in} \sim O(M_{Pl})$ then both eqs. \rf{c.10} and
\rf{c.19} give close limitations on $m_{\varphi}$. However, for
$\varphi_{in} \gtrless M_{Pl}$ we should use eq. \rf{c.10}, \rf{c.19}
correspondingly.

In the case of decaying gravexcitons the CMBR anisotropy due to 
gravexciton isocurvature fluctuations is washed out.


\bigskip
{\bf Acknowledgments}

We would like to thank
Valery Rubakov and Alexander Sakharov for valuable correspondence, 
Nemanja Kaloper
for useful comments concerning
his recent work \cite{KL} and Martin Rainer for interesting discussions.
A.Z. thanks H. Nicolai
and the Albert Einstein Institute for kind hospitality. U.G. acknowledges
financial support from DFG grant KON 1575/1999/GU 522/1.


%
%

\begin{thebibliography}{99}
%
\bibitem{Kolb}
E.W. Kolb, M.J. Perry and T.P. Walker, Phys. Rev. D33 (1986) 869 - 871.

%
\bibitem{iban}
L.E. Ib\'a\~nez, {\it
The second string (phenomenological) revolution},
\, hep-ph/9911499.

\bibitem{BBSMS}
T. Banks, M. Berkooz, S.H. Shenker,
G. Moore and P.J. Steinhardt, Phys. Rev. D52, (1995), 3548 - 3562,
\, hep-th/9503114.

\bibitem{RSh}
V.A. Rubakov and M.E. Shaposhnikov,
Phys. Lett. B125, (1983), 136-138;

\bibitem{Akama}
K. Akama, {\it Pregeometry}, (Lecture Notes in Physics, Gauge Theory
and Gravitation, Proc. Int. Symp. {\it Gauge Theory and Gravitation},
Nara , Japan, August 20-24, 1982), Springer-Verlag, 1983, Ed. K.Kikkawa,
N. Nakanishi and H. Nariai, p. 267 - 271.

\bibitem{Vi}
M. Visser, Phys. Lett. B159, (1985), 22 - 25.

\bibitem{HW}
P. Ho\v rava and E. Witten, Nucl. Phys. B460, (1996), 506, hep-th/9510209;
P. Ho\v rava and E. Witten, Nucl. Phys. B475, (1996), 94, hep-th/9603142.

\bibitem{W2}
E. Witten, Nucl. Phys. B471, (1996), 135, hep-th/9602070.

\bibitem{Ly1}
J.D. Lykken, Phys. Rev. D54, (1996), 3693, hep-th/9603133.

\bibitem{sub-mill1}
N. Arkani-Hamed, S. Dimopoulos and G. Dvali,
Phys. Lett. B429, (1998), 263 - 272, \, hep-ph/9803315.

\bibitem{sub-mill1a}
I. Antoniadis, N. Arkani-Hamed, S. Dimopoulos and G. Dvali,
Phys. Lett. B436, (1998), 257 - 263, \, hep-ph/9804398.

\bibitem{Tye}
G. Shiu and S.H. Tye,
Phys. Rev. D58, (1998), 106007,
\, hep-th/9809582.

\bibitem{Iban2}
L.E. Ib\'a\~nez, C. Mu\~noz and S. Rigolin,
Nucl. Phys. B553, (1999), 43 - 80 \, hep-ph/9812397.

%
\bibitem{sub-mill2}
N. Arkani-Hamed, S. Dimopoulos and G. J. March-Russell, {\it Stabilization
of sub-millimetre dimensions: the new guise of the hierarchy
problem},\, hep-th/9809124.

%
\bibitem{sub-mill3}
N. Arkani-Hamed, S. Dimopoulos, N. Kaloper and J. March-Russell,
{\it Rapid asymmetric inflation and early cosmology in theories
with sub-millimetre dimensions},\, hep-ph/9903224.

\bibitem{GZ(PRD)}
U. G\"unther and A. Zhuk, Phys. Rev. D56, (1997), 6391 - 6402,
\, gr-qc/9706050.

\bibitem{Beersh}
U. G\"unther and A. Zhuk,  {\it Stable compactification and
gravitational excitons from extra di\-men\-sions}, (Proc. Workshop
{\it Modern
Modified Theories of Gravitation and Cosmology}, Beer Sheva, Israel, June
29 - 30, 1997), Hadronic Journal 21, (1998),  279 - 318, \,
gr-qc/9710086.
%
\bibitem{GKZ}
U. G\"unther, S. Kriskiv and A. Zhuk, Gravitation
and Cosmology 4, (1998), 1 - 16, \,  gr-qc/9801013.

\bibitem{GZ(CQG)}
U. G\"unther and A.Zhuk, Class. Quant. Grav. 15,
(1998), 2025 - 2035, \, gr-qc/9804018.

%
\bibitem{RS}
L. Randall and R. Sundrum,
Phys. Rev. Lett. 83, (1999), 3370 - 3373, \, hep-ph/9905221;
Phys. Rev. Lett. 83, (1999), 4690 - 4693, \, hep-th/9906064.
%
%
\bibitem{ADDK}
N. Arkani-Hamed, S. Dimopoulos, G. Dvali and N. Kaloper, {\it Infinitely
large new dimensions},\, hep-th/9907209.
%
\bibitem{K2}
N. Kaloper, {\it
Crystal Manyfold Universes in $AdS$ Space},\, hep-th/9912125.
%
\bibitem{Ezawa1}
Y. Ezawa, Class. Quant. Grav. 6, (1989), 1267 - 1271.
%
\bibitem{Ezawa2}
Y. Ezawa, T. Watanabe and T. Yano, Progr. Theor. Phys. 86, (1991), 89 - 102.
%
%
\bibitem{KO1}
P. Kanti and K.A. Olive, Phys.Rev. D60, (1999), 043502 , hep-ph/9903524.
%
\bibitem{KO2}
P. Kanti and K.A. Olive, {\it Assisted chaotic inflation in higher
dimensional theories},\, hep-ph/9906331 .
\bibitem{KL}
N. Kaloper and A. Liddle, {\it Dynamics and perturbations in assisted
chaotic inflation},\, hep-ph/9910499.
%
\bibitem{York}
J.W. York, Phys. Rev. Lett. 28, (1972), 1082 - 1085.
%
\bibitem{GH}
G.W. Gibbons and S.W. Hawking, Phys. Rev. D15, (1977), 2752 - 2756.
%
\bibitem{GZ(non-lin)}
U. G\"unther and A. Zhuk, {\it Dynamics of non-linear multidimensional
cosmological models}\, (in preparation).
%
\bibitem{IMZ}
V.D. Ivashchuk, V.N. Melnikov and A.I. Zhuk, Nuovo Cimento B104,
(1989), 575 - 582.
%
\bibitem{RZ}
M. Rainer and A. Zhuk, Phys. Rev. D54, (1996), 6186 - 6192.
%
\bibitem{BZ}
U. Bleyer and A. Zhuk, Class. Quant. Grav. 12, (1995), 89 - 100.
%
%
\bibitem{Maz}
A. Mazumdar, Phys. Lett. B469, (1999), 55 - 60, \,hep-ph/9902381.
%
\bibitem{LL}
A.R. Liddle and D.H. Lyth, Phys. Rep. 231, (1993), 1 - 105.
%
\bibitem{LMS}
A. Liddle, A. Mazumdar and F. Schunck, Phys. Rev. D58, (1998), 061301, \,
astro-ph/9804177.
%
\bibitem{CFKRR}
G. Coughlan, W. Fischler, E. Kolb, S. Raby and G. Ross, Phys. Lett. B131,
(1983), 59 - 64.
%
\bibitem{ENQ}
J. Ellis, D. Nanopoulos and M. Quiros, Phys. Lett. B174, (1986), 176 - 182.
%
\bibitem{CCQR}
B. de Carlos, J. Casas, F. Quevedo and E. Roulet, Phys. Lett. B318, (1993),
447 - 456.
%
\bibitem{BKN}
T. Banks, D. Kaplan and A. Nelson, Phys. Rev. D49, (1994), 779 -787.
%
\bibitem{BBS}
T. Banks, M. Berkooz and P. Steinhardt, Phys. Rev. D52, (1995), 705 - 716,\,
hep-th/9501053.
%
%
\bibitem{KT}
E.W. Kolb and M.S. Turner, {\it The Early Universe}, Addison-Wesley
Publishing
Company, NY 1994.
%
\bibitem{GSZ}
U. G\"unther, A. Starobinsky and A. Zhuk, {\it Interacting gravitational
excitons from extra dimensions}, \, (in preparation).
%
\bibitem{PWW}
J. Preskill, M. Wise and F. Wilczek, Phys. Lett. B120, (1983), 127 - 132.
%
\bibitem{GLV}
A. Goncharov, A. Linde and M. Vysotsky, Phys. Lett. B147, (1984), 279 - 283.
%
\bibitem{PTW}
S. Perlmutter, M.S. Turner and M. White, Phys. Rev. Lett. 83, (1999), 670 -
673, \, astro-ph/9901052.
%

%
\bibitem{GZ(nl2)}
U. G\"unther and A. Zhuk, {\it Non-linear multidimensional cosmological
models with forms: the cosmological constant problem, stable
compactification and inflation problems} (in preparation).
%
\end{thebibliography}
\end{document}